\documentclass[10pt,letterpaper,american]{article}
\usepackage[latin9]{inputenc}
\usepackage{color}
\usepackage{babel}
\usepackage{units}
\usepackage{amsmath}
\usepackage{amssymb}
\usepackage{graphicx}
\usepackage[unicode=true,
 bookmarks=false,
 breaklinks=false,pdfborder={0 0 1},backref=false,colorlinks=true]
 {hyperref}
\hypersetup{
 citecolor=blue,urlcolor=blue}

\makeatletter

\pdfpageheight\paperheight
\pdfpagewidth\paperwidth

\usepackage{opticameet3}
\usepackage{capt-of}
\usepackage{booktabs}

\usepackage{cite}
\usepackage[printonlyused]{acronym}
\acrodef{DM}{distribution matcher}
\acrodef{Hi-DM}{hierarchical DM}
\acrodef{PS}{probabilistic shaping}
\acrodef{PAS}{probabilistic amplitude shaping}
\acrodef{AWGN}{additive white Gaussian noise}
\acrodef{QAM}{quadrature amplitude modulation}
\acrodef{FEC}{forward error correction}
\acrodef{PAM}{pulse amplitude modulation}
\acrodef{LUT}{look up table}
\acrodef{MB}{Maxwell-Boltzmann}
\acrodef{ESS}{enumerative sphere shaping}
\acrodef{CCDM}{constant composition distribution matching}
\acrodef{IR}{information rate}
\acrodef{SNR}{signal to noise ratio}
\acrodef{PPM}{pulse position modulation}
\acrodef{invDM}{inverse DM}
\acrodef{TX}{transmitter}
\acrodef{RX}{receiver}
\acrodef{BER}{bit error rate}
\acrodef{SER}{symbol error rate}

\makeatother

\begin{document}
\title{Practical Implementation of Sequence Selection for Nonlinear Probabilistic
Shaping}
\author{S. Civelli\textsuperscript{1,2,{*}}, E. Forestieri\textsuperscript{1,2},
M. Secondini\textsuperscript{1,2}}
\address{\textsuperscript{1} TeCIP Institute, Scuola Superiore Sant'Anna,
Via G. Moruzzi 1, 56124, Pisa, Italy\\
\textsuperscript{2} PNTLab, CNIT, Via G. Moruzzi 1, 56124, Pisa,
Italy}
\email{\textsuperscript{{*}}stella.civelli@santannapisa.it}

\maketitle
\copyrightyear{2022}
\begin{abstract}
We propose two novel techniques to implement sequence selection (SS)
for fiber nonlinearity mitigation, demonstrating a nonlinear shaping
gain of $0.24$\,bits/s/Hz, just $0.1$bits/s/Hz below the SS capacity
lower bound. 
\end{abstract}

\section{Introduction}

Probabilistic amplitude shaping (PAS) is employed in current high
capacity optical fiber communication systems to maximize the spectral
efficiency and provide rate flexibility \cite{bocherer2015bandwidth,buchali2016JLT}.
Besides its well studied advantages in the linear regime, PAS has
been investigated also for the opportunity to reduce nonlinear effects
\cite{fehenberger2016JLT,geller2016shaping,gultekin2021kurtosisESS,cho2021shaping}.
Recently, sequence selection (SS) has been proposed as a theoretical
approach to lower bound the capacity of the nonlinear fiber channel
\cite{civelli2021ecoc,secondini2022jlt}. In the SS approach, the
input distribution is optimized by selecting only ``good sequences''
(generating less nonlinear interference) for transmission. Unfortunately,
SS does not provide a practical way to map information bits on good
sequences. Therefore, a practical implementation of SS to be included
in PAS is missing, except for a preliminary attempt using constant
composition DM \cite{wu2022listCCDM}. In this work, we propose two
novel techniques to practically implement SS.

\section{Implementation of Sequence Selection}

Here we present two possible practical implementations of SS, namely
bit scrambling SS (BSSS) and symbol interleaving SS (SISS). In both
implementations, blocks of input bits are mapped to $N_{t}$ different
test sequences of $n$ dual-polarization QAM (4D) symbols. The sequences
are compared according to a proper metric and the best one is transmitted.
Pilot bits or symbols and $N_{t}-1$ different invertible transformations
are used to ensure that the test sequences behave as if they were
independently drawn from a given distribution (uniform or shaped),
while remaining univocally decodable. The two techniques differ for
the employed invertible transformation, for the concatenation order
of the latter with FEC, and for the use of pilot bits or symbols.
The performance of both techniques depends on the acceptance rate
$1/N_{t}$, on the rate loss due to pilot bits or symbols, and on
the accuracy of the selection metric. The choice of the metric, which
must be computed $N_{t}$ times, also impacts on the computational
complexity. While a thorough complexity analysis is deferred to a
future study, here we consider two extreme cases: the more accurate
and complex nonlinear interference (NLI) metric$||\mathbf{x}-\mathbf{y}||$,
where $\mathbf{x}$ is the transmitted sequence and $\mathbf{y}$
the corresponding sequence obtained through a numerically emulated
single-channel noiseless propagation of $\mathbf{x}$ \cite{secondini2022jlt};
and the extremely simple windowed Kurtosis (WK) \cite{cho2021shaping}.

\subsection{Bit Scrambling SS\label{subsec:RTSSbits}}

\begin{figure}
\centering \includegraphics[width=1\columnwidth]{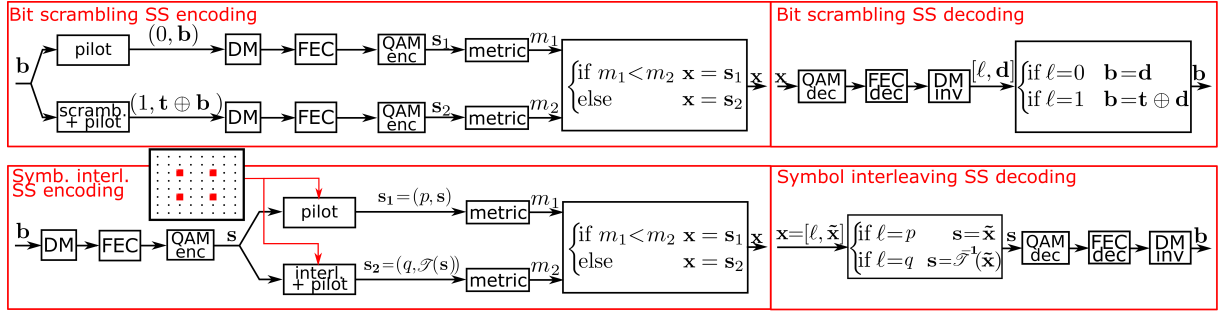}\caption{\label{fig:results}Bit scrambling and symbol interleaving SS , with
$N_{t}=2$ test sequences.}
\end{figure}

The working principle of BSSS is sketched at the top of Fig.~\ref{fig:results}
for $N_{t}=2$ test sequences. Given a sequence $\mathbf{b}$ of information
bits, the two test sequences are obtained by prepending the pilot
bit $0$ and $1$ to, respectively, $\mathbf{b}$ and $\mathbf{t}\oplus\mathbf{b}$,
where $\mathbf{t}$ is a random-like bit sequence (fixed and known
to the receiver), and $\oplus$ denotes the XOR operation. Each test
sequence, after going through the DM (if PAS is used) and the FEC
encoder, is mapped to a sequence of $n$ 4D symbols. The corresponding
metrics are then evaluated and the best sequence is transmitted. At
the receiver side, after dispersion compensation (and possibly other
processing blocks not shown here for simplicity), the received sequence
is processed by the QAM demapper, FEC decoder, and inverse DM. Finally,
the information bits are recovered by removing the pilot bit and,
if the latter is equal to $1$, by applying again the XOR operation
to descramble the bit sequence. The technique can be extended to the
case of $N_{t}>2$ test sequences by using $\left\lceil \log_{2}N_{t}\right\rceil $
pilot bits\footnote{$\lceil x\rceil$ denotes the smallest integer greater than or equal
to $x$.} and $N_{t}-1$ different scrambling sequences $\mathbf{t}_{1},\ldots,\mathbf{t}_{N_{t}-1}$.

BSSS is conceived to approximate the theoretical SS approach proposed
in \cite{civelli2021ecoc,secondini2022jlt} to lower-bound channel
capacity. It provides practically independent test sequences with
minimum rate loss, is compatible with the PAS approach with reverse
FEC concatenation, and ensures that the pilot bits are protected by
FEC. However, it poses a constrain on the length $n$ of the test
sequences (it must correspond to one or more FEC codewords) and requires
applying the DM and FEC encoder to each test sequence, with a subsequent
increase of complexity. To overcome these limitations, the alternative
SISS technique proposed below may be employed. 

\subsection{Symbol Interleaving SS \label{subsec:RTSSsymb}}

The working principle of SISS is sketched at the bottom of Fig.~\ref{fig:results}
for $N_{t}=2$ test sequences. In this case, the DM, FEC encoder,
and QAM mapper are directly applied to the information bit sequence
$\mathbf{b}$ to obtain the sequence of 4D symbols $\mathbf{s}$.
Then, the two test sequences are obtained by prepending two different
pilot symbols, $p$ and $q$, respectively to $\mathbf{s}$ and $\mathcal{T}(\mathbf{s})$,
where $\mathcal{T}(\cdot)$ denotes interleaving according to a randomly-selected
permutation table (fixed and known to the receiver). The random permutation
yields a sufficiently different new sequence (in terms of generated
NLI), but with the same amplitude distribution induced by PAS within
the block.  The corresponding metrics are then evaluated and the
best sequence is transmitted. At the receiver side, deinterleaving
$\mathcal{T}^{-1}(\cdot)$ is performed or not depending on the value
of the pilot symbol, followed by QAM demapping, FEC decoding, and
inverse DM. The technique can be extended to the case of $N_{t}>2$
test sequences by using one or more pilot symbols and $N_{t}-1$ different
interleaving maps. Pilot symbols, which are not protected by FEC,
are drawn from the dual-polarization QPSK constellation shown in the
inset of Fig.~\ref{fig:results}, properly carved out of the full
QAM constellation to ensure a sufficiently low error rate. With this
choice, one pilot symbol allows the identification of up to 16 different
interleaving maps, so that $\lceil\log_{2}(N_{t})/4\rceil$ pilot
symbols are used in total. In contrast with BSSS, the length of the
test sequences is not constrained by the length of the FEC codewords.
In fact, if desired, the symbol sequence $\mathbf{s}$ can be divided
into several shorter subsequences, which are then individually processed
by the remaining blocks of the scheme of Fig.~\ref{fig:results}.

\section{System Performance\label{sec:systemsetupperformance}}

The two implementations of SS are tested by means of numerical simulations
on a $30\times100$\,km single mode fiber link with EDFA ($\beta_{2}=\unit[21.7]{ps^{2}/km}$,
$\gamma=\unit[1.27]{W^{-1}m^{-1}},$ $\alpha_{\text{dB}}=\unit[0.2]{dB/km}$,
$N_{F}=\unit[5]{dB}$). Five $\unit[46.5]{GBd}$ WDM channels, spaced
$50$\,GHz apart, are propagated across the link. Each WDM channel
is obtained by modulating root-raised-cosine pulses with 0.05 roll-off
by the dual polarization PAS-64QAM symbols obtained according to the
schemes of Fig.~\ref{fig:results}. Enumerative sphere shaping (ESS)
\cite{gultekin2018Sphereshaping} with rate $R=1.3$\,bits/amplitude
(slightly increased to compensate for the rate loss due to pilot bits
or symbols when BSSS or SISS are employed) and blocklength 256 (optimized
for nonlinear performance) is used to implement PAS. BSSS and SISS
are implemented on sequences of $n=256$ 4D symbols, after PAS. FEC
encoding and decoding are omitted from the simulations. At the receiver
side, after channel demultiplexing, dispersion compensation, matched
filtering, symbol-time sampling, and compensation of the mean phase
rotation, the spectral efficiency (SE) corresponding to the achievable
information rate with bit-wise decoding is evaluated \cite{fehenberger2016JLT}.

\begin{figure}
\centering\includegraphics[width=0.85\columnwidth]{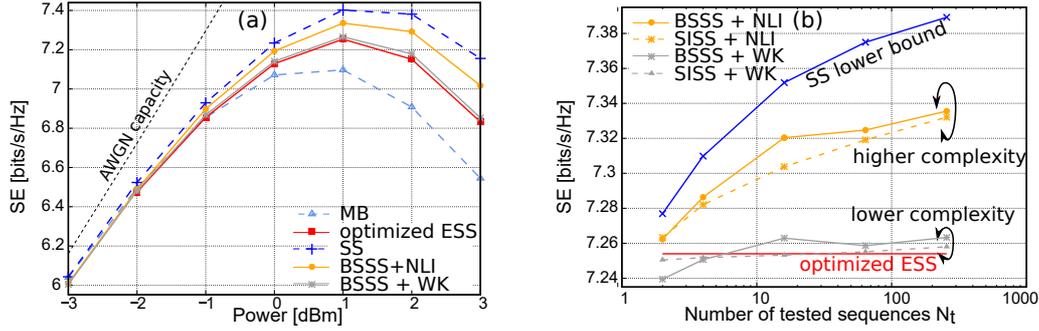}\caption{\label{fig:results-1}(a) SE with different shaping techniques; (b)
SE vs number of tested sequences $N_{t}$}
\end{figure}

Fig.~\ref{fig:results-1}(a) reports the SE obtained with the BSSS
scheme ($N_{t}=256$) and two different metrics; with i.i.d. Maxwell--Boltzmann
(MB) distributed symbols (optimal in the linear regime) \cite{kschischang1993optimal};
with ESS alone; and the theoretical SE achievable by the SS approach
(computed as explained in \cite{secondini2022jlt} with acceptance
rate $\eta=10^{-3}$ and averaged cost function). BSSS with the NLI
metric provides a gain of $0.24$ and $0.08$\,bits/s/Hz w.r.t. to
MB and ESS, respectively. On the other hand, BSSS with the WK metric
(which basically measures the intensity fluctuations within the block)
performs only slightly better than ESS. Eventually, the SS lower bound
shows that an additional gain of at least $0.1$\,bits/s/Hz is theoretically
achievable by improving the selection metric and increasing the number
of tested sequences. 

Next, Fig.~\ref{fig:results-1}(b) shows the performance at the optimal
power when changing the number of tested sequences $N_{t}$, for both
BSSS and SISS and for both the NLI and WK metrics. The SS lower bound
(with same acceptance rate $1/N_{t}$ and averaged cost function \cite{secondini2022jlt})
is also reported as a reference. In general, the performance improves
when increasing the number of tested sequences $N_{t}$. The gain
obtained with SISS is slightly below the one obtained with BSSS, mainly
because of the larger rate loss due to the use of pilot symbols compared
to pilot bits. The complex NLI metric is always superior to the simple
WK metric, while the theoretical bound shows that higher gains can
be achieved by further improving the selection metric.

\section{Conclusion}

Two practical implementations of SS have been proposed. The two techniques,
which differ in the concatenation order with FEC and PAS, provide
a similar nonlinear shaping gain, outperforming linearly optimized
PAS by 0.24~bit/s/Hz, and optimal-blocklength (for the nonlinear
regime) ESS by 0.08 bit/s/Hz. A variable acceptance rate and two different
selection metrics have been considered, showing the possibility to
reduce complexity by sacrificing performance, and highlighting the
importance of finding an accurate metric with sufficiently low complexity.

\section*{Acknowledgment}

This work was supported by Huawei.

\bibliographystyle{opticajnl}

\end{document}